\documentclass[twocolumn]{IEEEtran}
\usepackage{graphicx}
\usepackage{epstopdf}
\graphicspath{{Images/}}
\DeclareGraphicsExtensions{.eps,.jpeg,.png}

\usepackage[cmex10]{amsmath}
\usepackage{amsthm}
\usepackage{amsfonts}
\usepackage{cite}
\usepackage{subfig}
\usepackage[framemethod=default]{mdframed}
\usepackage{showexpl}
\mdfdefinestyle{exampledefault}{%
linecolor=red, innerleftmargin=5,innerrightmargin=5,
frametitlerule=true,frametitlerulecolor=green,
frametitlebackgroundcolor=yellow,
frametitlerulewidth=2pt}

\newcommand\numberthis{\addtocounter{equation}{1}\tag{\theequation}}

\usepackage{algorithm}
\usepackage{algpseudocode}
\usepackage{pifont}
\usepackage{fixltx2e}
\usepackage{caption}
\usepackage{soul}

\usepackage{tikz}

\newcommand\encircle[1]{%
  \tikz[baseline=(X.text)] 
    \node (X) [draw, shape=circle, inner sep=0, thick] {\strut #1};}

\DeclareRobustCommand\mytikzdot{\encircle}

\begin{document}
\title{The New Frontier in RAN Heterogeneity: Multi-Tier Drone-Cells}
%
%
%
\author{
Irem~Bor-Yaliniz,~
and~Halim~Yanikomeroglu
\thanks{------------------------------------------------------------------}
\thanks{I. Bor-Yaliniz and H. Yanikomeroglu are with the Department of Systems and Computer Engineering, Carleton University, Ottawa, Ontario, Canada. (e-mail: \{irembor, halim\}@sce.carleton.ca).} 
}

%



\maketitle
\begin{abstract}
%
In cellular networks, the locations of the radio access network (RAN) elements are determined mainly based on the long-term traffic behaviour. However, when the random and hard-to-predict spatio-temporal distribution of the traffic (load, demand) does not fully match the fixed locations of the RAN elements (supply), some performance degradation becomes inevitable. The concept of multi-tier cells (heterogeneous networks, HetNets) has been introduced in 4G networks to alleviate this mismatch. However, as the traffic distribution deviates more and more from the long-term average, even the HetNet architecture will have difficulty in coping with the erratic supply-demand mismatch, unless the RAN is grossly over-engineered (which is a financially non-viable solution). In this article, we study the opportunistic utilization of low-altitude unmanned aerial platforms equipped with base stations (BSs), i.e., \textit{drone-BSs}, in future wireless networks. In particular, we envisage a \textit{multi-tier drone-cell} network complementing the terrestrial HetNets. The variety of equipment and non-rigid placement options allow utilizing multi-tier drone-cell networks to serve diversified demands. Hence, drone-cells bring the supply to where the demand is, which sets new frontiers for the heterogeneity in 5G networks. We investigate the advancements promised by drone-cells and discuss the challenges associated with their operation and management. We propose a drone-cell management framework (DMF) benefiting from the synergy among software-defined networking (SDN), network functions virtualization (NFV), and cloud computing. We demonstrate DMF mechanisms via a case study, and numerically show that it can reduce the cost of utilizing drone-cells in multi-tenancy cellular networks. 
%
\end{abstract}


%
\IEEEpeerreviewmaketitle

\section{Introduction}
%
Transportation and communication technologies are major contributors to our lifestyles. Combining the state-of-the-art advancements in these two technologies, drone-assisted mobile communications has gained momentum rapidly. Drones equipped with transceivers, i.e., drone base stations (\textit{drone-BSs}) forming \textit{drone-cells}, can help satisfy the demands of the future wireless networks~\cite{bor_2016}.\footnote{Drone connectivity scenarios in recent 3GPP Release 14 documents (e.g., 3GPP TR 22.862 V14.0.0 (2016-06)) only include remote control of drones, which is different to the vision of drone-cells. Also, considering the limited time remaining until the development of 5G standards, we envision that drone-BSs can be utilized in beyond-5G/6G wireless networks (rather than 5G).} Moreover, they can utilize the latest radio access technologies (RATs), such as millimeter wave (mmWave), and free-space optical communication (FSO). Miscellaneous assets of drones and placement options provide opportunity to create \textit{multi-tier drone-cell networks} to enhance connectivity whenever, wherever, and however needed. Therefore, the main advantage of drone-cells is the radical flexibility they create.

The phenomenon of providing ubiquitous connectivity to diversified user and device types is the key challenge for 5G and beyond-5G wireless networks. The Achilles' heel of the proposed technologies, such as decreasing cell size, cloud radio access networks (C-RAN), distributed antenna systems (DAS), and heterogeneous network (HetNet) deployments, is their rather rigid design based on long-term traffic behaviour~\cite{demestichas_5g_2013}. In case of unexpected and temporary events creating hard-to-predict inhomogeneous traffic demand~\cite{hethetnets}, such as natural disasters, traffic congestions, or concerts, wireless networks may need additional support to maintain ubiquitous connections. Drone-cells address this need by increasing relevance between the distributions of supply (BSs) and demand (user traffic). They can be used opportunistically to leverage the heterogeneity, i.e., by dynamically deploying BSs with different power levels and RATs.


Although discussions on utilizing drone-cells in cellular networks have flourished recently~\cite{bor_2016, hourani_final}, the readiness of cellular networks to employ such dynamic nodes has not been discussed. For instance, drone-cells require seamless integration to the network during their activity and seamless disintegration when their service duration is over. This requires the capability of configuring the network efficiently, for which configuration and management  flexibilities, and self-organizing capabilities of the 3GPP Long-Term Evolution (LTE) networks may not be adequate. Hence, updating the network, such as for adding new applications, tools, and technologies, is time and money consuming~\cite{bradai_cellular_2015}. Also, massive amounts of granular information about users and networks must be continuously collected and analysed by intelligent algorithms. Collecting, storing, and processing big data is challenging for existing wireless networks~\cite{demestichas_5g_2013}. Moreover,  it is not yet clear how to balance centralized (e.g., mobile cloud) and distributed (e.g., mobile edge computing) paradigms~\cite{bradai_cellular_2015}.

Recent proposals for future wireless network architectures aim for creating a flexible network with improved agility and resilience. Cloud computing, software-defined networking (SDN), and network functions virtualization (NFV) have been proposed to relax the entrenched structure of the wireless networks, increase openness, ease configuration, and utilize cloud computing for storing and analysing big data. At the same time, these technologies may decouple the roles in the business model into infrastructure providers (InPs), mobile virtual network operators (MVNOs), and service providers (SPs)~\cite{liang_wireless_2015}, which also changes the owners and sources of information. 

In order to utilize drone-cells in future wireless networks, we propose a drone-cell management framework (DMF) and discuss the related business and information models. The proposed framework relies on creating intelligence from big data in the cloud and re-configuring the network accordingly by SDN and NFV. In the following section, we describe the drone-cells, the motivations for utilizing them in wireless networks, and the challenges. Then we introduce DMF, discuss business and information models, and challenges. Finally, we demonstrate the fundamental principles of DMF via a case study; the Conclusion section closes the paper. 
\section{Descriptions, Opportunities, and Challenges}
\label{sec:description}
\begin{figure*}[!t]
\centering
\includegraphics[width=1\textwidth]{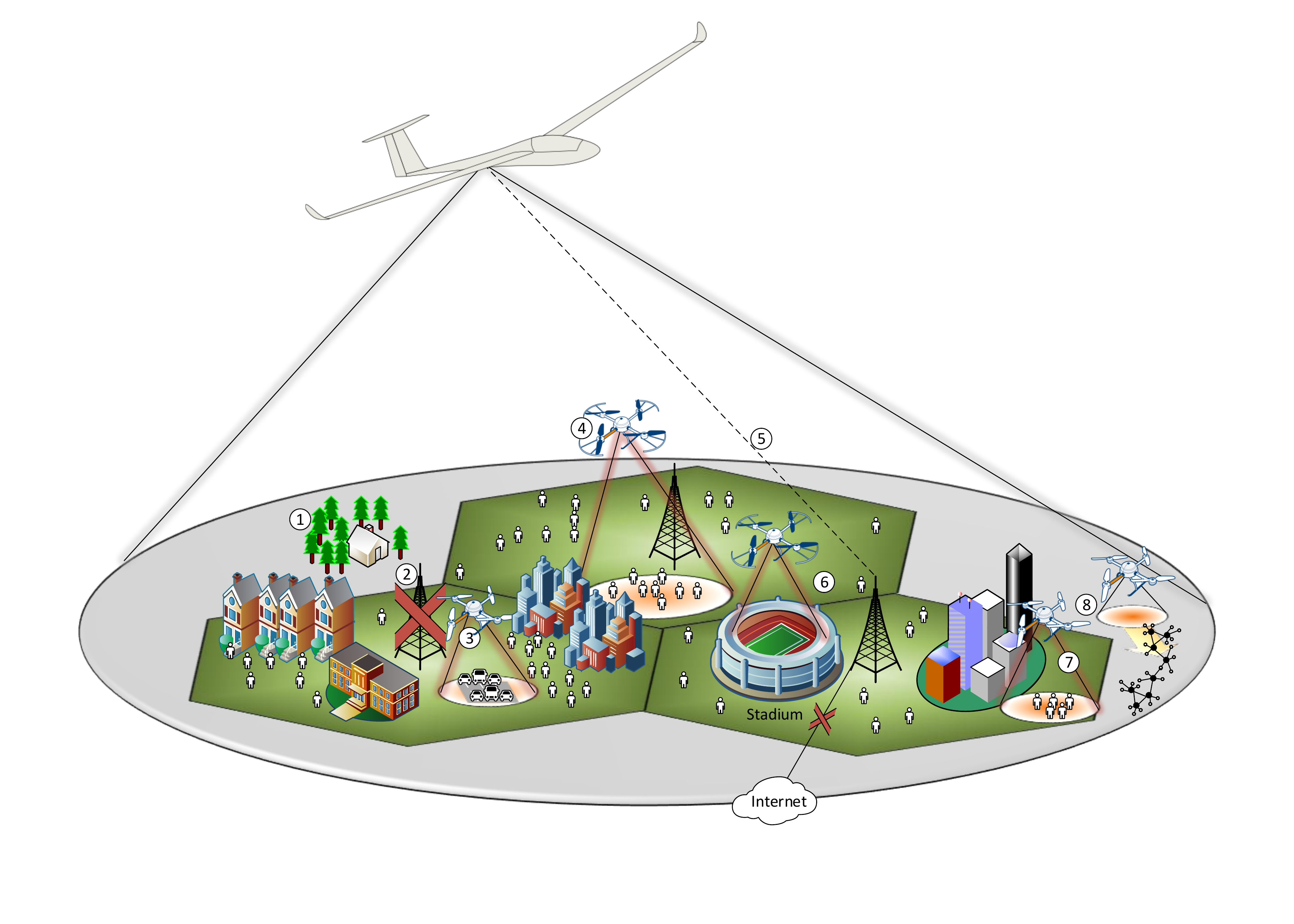}
\captionsetup{font=small}
\caption{Multi-tier drone-cell networks can be used for many scenarios: \mytikzdot{1} Providing service to rural areas (macro-drone-cell), \mytikzdot{2} Deputizing for a malfunctioning BS (macro-drone-cell), \mytikzdot{3} Serving users with high mobility (femto-drone-cell), \mytikzdot{4} Assisting a macrocell in case of RAN congestion (pico-drone-cell), \mytikzdot{5} Assisting a macrocell in case of core network congestion or malfunctioning (macro-drone-cell), \mytikzdot{6} Providing additional resources for temporary events, e.g., concerts and sports events, \mytikzdot{7} Providing coverage for temporary blind spots, and \mytikzdot{8} Reducing energy dissipation of sensor networks by moving towards them (femto-drone-cell).}
\label{fig:multi_tier}
\end{figure*}

A drone-BS is a low-altitude\footnote{The classification of drones is a rather involved task due to their variety~\cite[Ch. 5]{handbook}. However, in this context, the term ``low-altitude" is used to differentiate the drone-BSs from the high altitude platforms (HAPs) operating over 20 km.} unmanned aerial vehicle equipped with transceivers to assist the wireless networks~\cite{bor_2016}, and \textit{drone-cell} is the corresponding coverage area. Size of a drone-cell varies based on the drone-BS's altitude, location, transmission power, RATs, antenna directivity, type of drone, and the characteristics of the environment. Hence, multi-tier drone-cell networks can be constructed by utilizing several drone types, which is similar to the terrestrial HetNets with macro-, small-, femtocells, and relays . A multi-tier drone-cell network architecture, assisting the terrestrial HetNets in several cases, is depicted in Fig.~\ref{fig:multi_tier}. 

%
Drone-cells are useful in scenarios requiring agility and resiliency of wireless networks because they can prevent over-engineering. These type of scenarios can be categorized as \textit{temporary}, \textit{unexpected}, and \textit{critical}, as shown in Table~\ref{tab:metis}, where relevant test cases of the METIS\footnote{Mobile and Wireless Communications Enablers for Twenty-twenty (2020) Information Society.} project are listed~\cite{metis}. Based on the scenario, the benefit to the network from a drone-cell varies. For instance, in traffic jam, stadium, and dense urban information society scenarios, a drone-cell can help prevent unexpected or temporary congestion in the network. Alternatively, drone-cells can improve resilience of wireless networks by providing additional coverage in case of a natural disaster, or by enabling teleprotection for the smart grid. 
\begin{table*}[!t]
\caption{An example of \textit{categorization of test cases of METIS requiring agility and resilience:} An event can fall under one category or multiple categories and each combination may require different solutions. For instance, connectivity requirements in case of an only temporary event (e.g., stadium) may be addressed by over-engineering. Then, expenses of drone-BS operations may be compared to the expenses of over-engineering, including energy and maintenance costs. On the other hand, for both temporary and unexpected events, (e.g. traffic jam), drone-BSs may be utilized opportunistically. For temporary, unexpected and critical operations (e.g., emergency communications) drone-cells can provide much more than revenue, such as saving lives.}
\centering
\begin{tabular}{|c|c|c|c|}
\hline
\textbf{Test Case} & \textbf{Temporary} & \textbf{Unexpected} & \textbf{Critical}\\
\hline
 Stadium & X &  & \\
\hline
Teleprotection in smart grid &  & X & \\
\hline
Traffic jam & X & X & \\
\hline
Blind spots & X & X & \\
\hline
Open air festival & X &  & \\
\hline
Emergency communications & X & X & X\\
\hline
Traffic efficiency and safety &  &  & X\\
\hline
Dense urban information society &  X &  &  \\
\hline
Massive deployment of sensor-type devices & X &  X & X\\
\hline
\label{tab:metis}
\end{tabular}
\end{table*}

Critical scenarios have challenging demands, such as very high data rates, high reliability, or low energy consumption. Beyond the benefits to the network, providing connectivity in some of these scenarios is important to prevent serious losses, for example by saving lives in emergency communications, or increasing the lifetime of sensors and actuators at hard to reach areas. In case of emergency communications, and tele-control applications, drone-cells can enable high data rates and reliability, especially for situations in which the conventional modes of wireless access are either not present or difficult to establish. Mobility of drone-cells enables them to serve users with high mobility and data rate demand, e.g., for traffic efficiency and safety~\cite{metis}. Alternatively, sensor-type devices requiring low energy consumption can benefit from drone-cells. Instead of forcing low-power devices to transmit to farther BSs, or deploying small cells densely, mobile sinks can be used. A drone-cell can move towards clusters of devices and provide low-power communication due to its proximity and potential line-of-sight (LOS) connectivity. In particular, when unexpected events trigger massive sensor activity, drone-cells can reduce the overall stress on the network and increase the life-time of sensors. Note that the critical scenarios, in which the conventional wireless access options are not feasible, may render them as the first applications of drone-cells in providing (almost) carrier-grade service.  

Although the flexibility of drone-cells allows utilizing them in versatile scenarios, it creates significant design, operation, and management challenges, which are discussed next. 

\subsection{Challenges of drone-cells}
\label{sec:chal1}
\subsubsection{Efficient design} Drones have been utilized for military, surveillance and reconnaissance applications for long time. However, their usage in cellular communications as drone-BSs is a novel concept under investigation. For instance, a preliminary implementation of an LTE eNodeB-based drone operation is presented in~\cite{hourani_final}, where a remote radio head (RRH) is deployed on an off-the-shelf helikite. The helikite is tethered to a truck carrying the baseband unit (BBU), and optical fiber is used for the fronthaul. This tethered helikite design is due to the non-existence of drones that are specifically designed to operate as drone-BSs. 
Drones are generally designed for their task, which is the reason for their great variety~\cite[Ch. 5]{handbook}. 

Drone-BSs would have unique requirements that can benefit from special-purpose designs, such as long-time hovering, long endurance, robustness against turbulence, minimum wing-span allowing MIMO, and provision of energy for transmission (in addition to flying). For instance, a hybrid-drone can be designed with vertical take-off capability of rotorcrafts and with collapsible wings (equipped with MIMO antenna elements and solar panels for energy harvesting), which can be unfolded for efficient gliding. 

Designing the payload of drone-BSs is as important as determining their mechanics, e.g., size, aerodynamics, and maximum take-off weight~\cite[Ch. 9]{handbook}. For efficient usage of the limited volume, weight, and energy of drone-BSs, the payload can vary according to the scenario. Several possible drone-cell configurations are listed below:
\begin{itemize}
\item Drone-relay (\textit{``Drolay"}): Compared to small- or macro-BSs, relays require less processing power, because their RRH may be relatively simple and they may not require an on-board BBU. Hence, they operate with light payloads and potentially consume less power. The size and weight of RAN nodes may not be critical for terrestrial HetNets, however a lighter payload improves endurance and decreases CAPEX and OPEX\footnote{Capital expenditure (CAPEX) and operational expenditure (OPEX).} significantly in drone-cell operations.
\item Small-drone-BS: They resemble terrestrial small-BSs with wireless backhaul. If a reliable wireless fronthaul can be maintained despite the mobility of drone-BSs, its advantage is twofold: First, it alleviates the weight and processing power required for an on-board BBU. Second, if combined with C-RAN, it can allow cooperation. C-RAN is useful particularly for dense HetNets~\cite{demestichas_5g_2013}, or when a fleet of drone-BSs are deployed. Scenarios \encircle{3}, \encircle{4}, \encircle{7}, and \encircle{8} in Fig.~\ref{eq:big_pic} exemplify potential usage. 
\item Macro-drone-BS: They resemble terrestrial macro-BSs with wireless backhaul. They can be deployed for longer endurance, broader coverage, or increased reliability of the network, e.g., \encircle{1}, \encircle{5} and \encircle{6} (Fig.~\ref{eq:big_pic}). BBU can be included if a reliable wireless backhaul exists. Since coverage is strongly related to altitude and power, macro-drone-BSs may have a larger size, which allows more payload, e.g. medium-altitude long-endurance drones~\cite[Ch. 113]{handbook}.
\end{itemize}

In addition to the discussion above, efficient drone-cell design can be enhanced by advancements on low-cost and light-weight energy harvesting, high-efficiency power amplifiers, beyond visual LOS operations, and alternative fuels, to name a few.

\subsubsection{Backhaul/fronthaul connection} In terrestrial networks, wireless backhaul/fronthaul is considered when fiber connectivity is unaffordable, e.g., dense HetNets or rural BSs. However, it is inevitable for multi-tier drone-cell networks. FSO and mmWave are promising for their high-rate and low spectrum cost. However, their reliability and coverage are limited, especially for inclement weather conditions \cite{kaushal_free_2015, siddique_wireless_2015}. Although mobility of drone-cells help maintain LOS, it necessitates robustness against rapid channel variations.


\subsubsection{Placement} Terrestrial BSs are deployed based on long-term traffic behaviour and over-engineering when necessary. However, drone-cells require quick and efficient placement. Therefore, it is of critical importance to determine the parameters affecting a drone-cell's performance, such as its altitude, location, and trajectory based on the network demands~\cite{bor_2016, elham}. For instance, if a drone-cell is utilized to release congestion in RAN within a congested cell, the target benefit is to offload as many users as needed to the drone-cell~\cite{bor_2016}. Particularly, if the congestion is at the cell edge, the drone-cell can be placed right on top of the users there. On the other hand, if the congestion is at the backhaul, some of the most popular contents can be cached in a drone-cell for \textit{content-centric placement} (Sec.~\ref{sec:virt}). Moreover, placement of multi-tier drone-cell networks requires integrated evaluation of many other challenges.
%
%
%
%
%

\subsection{Challenges of multi-tier drone-cell networks}
\label{sec:chal2}
There are additional challenges of multi-tier drone-cell networks. Although these challenges are similar to those of terrestrial HetNets, the particular details related to drone-cells are discussed here. 
\begin{itemize}
\item \textbf{Physical layer signal processing:} The link between the drone-cell and terrestrial nodes, i.e., air-to-ground links, have different characteristics than terrestrial channels~\cite{bor_2016, willink_measurement_2015}. However, the research on air-to-ground links is not mature and the proposed channel models vary depending on factors such as temperature, wind, foliage, near-sea environments, urban environments, and the aircraft used for measurement campaigns, to name a few. For instance, higher ground speed causes rapid variation of spatial diversity; users at different locations with respect to the drone-BS can have different channel characteristics simultaneously~\cite{willink_measurement_2015}. Therefore, designing robust signaling mechanisms with strict energy constraints of drone-BSs is challenging.  
\item \textbf{Interference dynamics:} Drone-cells in proximity can suffer from co-channel interference for their air-to-ground links, and backhaul/fronthaul.  Moreover, a drone-cell's mobility creates Doppler shift, which causes severe inter-carrier interference for RATs at high frequencies (e.g., mmWave). In HetNets, interference of terrestrial and air-to-ground-channels can decrease capacity. Therefore, advanced interference management schemes, which consider the characteristics of air-to-ground links and mobility of drone-cells, are required.
\item \textbf{Cooperation among drone-cells:} The dynamic nature of multi-tier drone-cell networks requires cooperation among drone-cells for efficiency in radio resource management. In addition to that, drone-cells can cooperate to adapt to the mobility of the users to decrease handover, optimize power and resource allocations, and avoid collisions.
\item \textbf{Infrastructure decision and planning:} The number and assets of drone-cells (e.g., access technology, memory, and speed) to be utilized for a multi-tier drone-cell network depend on circumstances, such as inclement weather conditions, size of the area to be served, type of service (e.g., virtual reality, internet-of-things), target benefit of the network (e.g., congestion release, resilience, low-latency), or service duration. Also, utilizing drone-cells with different access technologies can reduce interference, and increase capacity of multi-tier drone-cell networks, e.g., utilizing a macro-drone-cell with RF and small-drone-cells with mmWave to prevent frequency re-use. Hence, InPs must have a fleet which can respond to possible scenarios. To optimize the fleet and construct an efficient network, information sharing among all parties of the network, i.e., InPs, MVNOs and SPs, is required.
\end{itemize}

Cost, lack of regulations, security, and airworthiness are among other challenges of drones. The vital point of the matter is considering the effects of utilizing drones in highly sophisticated cellular communication networks, rather than using them for stand-alone applications, e.g., aerial photography or inspection. Therefore, drone-cells require an equivalently sophisticated management system, which is discussed next. 

\section{The drone-cell management framework}
\label{sec:DMF}
A drone-cell is not a one-size-fits-all solution, instead, it is tailored based on the target benefit. Along with the management of individual drone-cells, multi-tier drone-cell networks require active organization and monitoring, e.g., for nodes changing location or cells becoming congested. Three capabilities are required to integrate drone-cells with already sophisticated cellular networks:%

\begin{itemize}
\item \textbf{Global information:} The information gathered by BSs alone may be inadequate to generate intelligence for managing drone-cells. Global information, including location, type, and habits of the users, functionality of the BSs, and the contents to deliver must be stored and analyzed centrally. Big data and cloud computing can be effective solutions for that purpose.
\item \textbf{Programmability:} Both drone-cells and network tools need to be programmed based on the network updates. Moreover, sharing the resources made available by a drone-cell can reduce the CAPEX and OPEX. NFV can provide these capabilities to the wireless networks.
\item \textbf{Control:} Wireless networks must be configured efficiently for seamless integration/disintegration of drone-cells, such as changing protocols and creating new paths. SDN can be useful to update the network automatically via a software-based control plane. 
\end{itemize}
The current LTE architecture does not embody all of these abilities, but cloud, SDN, and NFV technologies can enable a more capable wireless communication system~\cite{demestichas_5g_2013}. 
 
\subsection{Enabling Technologies for DMF}
\label{sec:enable}
In this subsection, we briefly explain the technologies that increase capabilities of wireless networks and the interactions that are required to efficiently manage drone-cell-assisted wireless communications. 
\subsubsection{Cloud and Big Data}
There are many ways to approach the problem of collecting and processing sufficient data (Table~II) in a timely manner for efficiently utilizing drone-cells. A cloud for drone-cells, consisting of computing power and data storage (Fig.~2), combined with big data analysis tools, can provide efficient and economic use of centralized resources for network-wide monitoring and decision making~\cite{bradai_cellular_2015, zhou_toward_2014}. If drone-cells are owned by a traditional mobile network operator (MNO) (Fig.~2), the cloud is merely the data center of the MNO (similar to a private cloud), where the MNO as an administrator can choose to share its knowledge with some other players or use it for its own business purposes. Alternatively, if the drone-BSs are provided by an InP, the InP can use the cloud to collect information from MVNOs and SPs (Fig.~2 and Table~II). In this case, it is particularly important to guarantee security, latency, and privacy. Benefit of the cloud can be better exploited with a programmable (softwarized) network allowing dynamic updates based on big data processing, for which NFV and SDN can be enabling technologies.

\subsubsection{Network Functions Virtualization}
NFV alleviates the need for deploying specific network devices (such as packet and serving gateways, deep packet inspection modules, and firewalls) for the integration of drone-cells~\cite{bradai_cellular_2015}. By virtualizing the above-network functions on general purpose servers, standard storage devices, and switches, NFV allows a programmable network structure, which is particularly useful for drone-cells requiring seamless integration to the existing network (\encircle{4} in Fig. 2). Furthermore, virtualization of drone-cells as shared resources among M(V)NOs can decrease OPEX for each party (Section~\ref{sec:virt})~\cite{liang_wireless_2015}. However, the control and interconnection of VNFs becomes complicated, for which SDN can be useful~\cite{bradai_cellular_2015}.
\subsubsection{Software Defined Networking}
By isolating the control and data planes of network devices, SDN provides centralized control, global view of the network, easy reconfiguration, and orchestration of VNFs via flow-based networking (\encircle{4} in Fig. 2). Specifically for cellular networks, a centralized SDN controller can enable efficient radio resource and mobility management~\cite{bradai_cellular_2015}, which is particularly important to exploit drone-cells. For instance, SDN-based load balancing proposed in~\cite{bradai_cellular_2015} can be useful for multi-tier drone-cell networks, such that the load of each drone-BS and terrestrial-BS is optimized precisely. An SDN controller can update routing such that the burst of traffic from the drone-cells is carried through the network without any bottlenecks~\cite{zhou_toward_2014}. Similarly, in case of a natural disaster that causes the network to partially malfunction, network health information in the cloud can be utilized via SDN to route the traffic of drone-cells through undamaged parts of the network. Because SDN allows updating switches simultaneously (e.g., for new forwarding rules), it allows faster switching between RATs~\cite{yazici}, which eases utilizing different RATs in multi-tier drone-cell networks. Furthermore, the architecture based on hierarchical SDN controllers for unified handoff and routing proposed in~\cite{yazici} can allow granular management of flows through drone-cells. For instance, the handoff strategy can be changed to a more complex proactive handoff for decreasing the latency of flows from drone-cells. Alternatively, DMF may collaborate with the mobility management entities for efficiency, e.g., a drone-cell can follow high-mobility users on a highway (\encircle{3} in Fig.~\ref{fig:multi_tier}) to reduce handover. For further exploitation for the new degree-of-freedom introduced by the mobility of the drone-cells, the footprint of the drone-cells can be adjusted to optimize paging and polling, and location management parameters can be updated dynamically via the unified protocols of SDN.

\subsection{Business and Information Models of DMF}
\label{sec:bus_and_info}
In traditional cellular networks, an MNO owns almost the entire cellular network, such as BSs and core network, and sharing among MNOs is limited. However, future cellular networks may be partitioned between InPs, MVNOs and SPs~\cite{liang_wireless_2015}. For instance, high sophistication of drone operations may result in the drone-cell operator becoming a separate business entity, such as a drone-InP. 

\begin{figure*}[!t]
\centering
\includegraphics[width=1\textwidth]{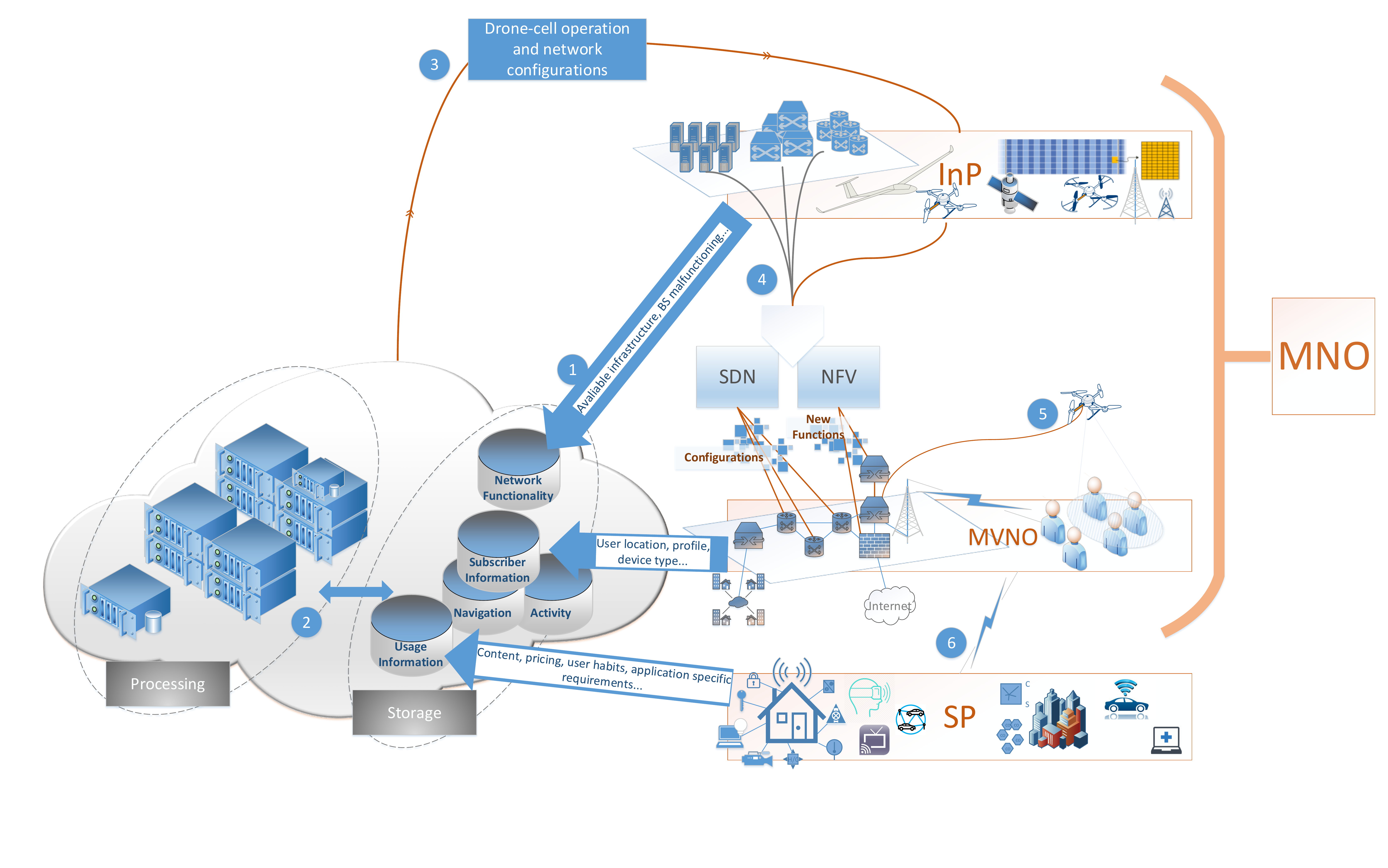}
\captionsetup{font=small}
\caption{DMF mechanism and potential business and information model demonstrating partitioning of the traditional MNO into InP (cloud, server, drone-BS etc.) and MVNO: \mytikzdot{1} Collect and store global data; \mytikzdot{2} Process data for network monitoring and creating intelligence; \mytikzdot{3} Provide guidance for drone-cell's operation (placement, content to be loaded, access technology, service duration, coverage area, moving patterns); \mytikzdot{4} Re-configure the virtual network of MVNO for drone-cell integration by SDN and NFV technologies, e.g., introduce another gateway to handle busy traffic and create new paths among the new and existing functions; \mytikzdot{5} Drone-cell assists the network; \mytikzdot{6} SP can continue delivering services successfully.}
\label{fig:model}
\end{figure*}

Fig.~\ref{fig:model} represents a DMF with potential business and information models, and shows what is owned by these parties, and what information flows from them to the cloud. According to the model, all physical resources of the cellular network, including drone-cells, BSs, spectrum, and core network, are owned by InPs. The MVNO is responsible for operating the virtual network efficiently such that the services of the SP are delivered to the users successfully. Note that, in this model, perfect isolation and slicing is assumed such that an MVNO has a complete virtual cellular network~\cite{liang_wireless_2015}. 

Compared to the traditional cellular networks, more granular data is available, but it is distributed unless collected in a cloud. A brief list of information, which can be critical for the operation of the DMF, is provided in Table~\ref{tab:info} along with its type, source, and usage~\cite{bradai_cellular_2015}. The results of the processing are then used to orchestrate SDN and NFV for the purpose of integrating drone-cells into the networks. This mechanism is demonstrated in Section~\ref{sec:virt}. 

Note that such isolated business roles may not be realistic for the near future. Instead, the role of an MNO may get partitioned into three actors, namely, InP, MVNO, and SP. Since it will mature in the long run, this partitioning should not be considered as siloing, but rather specialization. Accordingly, unique pricing strategies and QoS monitoring requirements will likely appear for drone-cell operations. Although complex and expensive, drone-cell operations can increase revenues in several ways, such as enabling a leaner terrestrial network, service to high-priority users (e.g., for public safety), and continuity of  challenging services even in cases of unpredictable high density traffic in areas with relatively insufficient infrastructure (Section~\ref{sec:description}).

%

%
\begin{table*}[!t]
\caption{Various information that can be gathered in the cloud.}
\label{tab:info}
\centering
\begin{tabular}{|c|c|c|c|}
\hline
\textbf{Information} & \textbf{Type} & \textbf{Source} & \textbf{Use}\\
\hline
International Mobile Subscriber Identity (IMSI) & User & MNO & True identity of the user\\
\hline
User profile information & User & MVNO & Subscription type, activities\\
\hline
User's location & Network & MVNO & Location\\
\hline
Device type & Network & MVNO & Location, resource allocation provisioning, etc.\\
\hline
Functionality of the nodes & Network & InP & Location, coverage extension, energy saving, etc.\\
\hline
User's activity and navigation & Network & MVNO & Placement, consumption, lifestyle, etc.\\
\hline
Content & Usage & SP & Centers of interest, preferences, pricing, content delivery, etc.\\
\hline
Long-term historic data & Usage & SP & Content delivery, pricing, etc.\\
\hline
\end{tabular}
\end{table*}

\subsection{Challenges for DMF Implementation}
\label{sec:challenges}
Network management required for DMF involves the challenges of NFV and SDN. Slicing of drone-cells, isolation of the traffic of different MVNOs, migration of virtual network functions, virtual resource management, and scheduling can be listed among the major challenges related to NFV~\cite{liang_wireless_2015}. Regarding the SDN in DMF, the main challenges are providing a global view to the SDN controller, i.e., scalability, efficiency in programming new paths, and communicating with different virtual network entities and application interfaces~\cite{sezer_are_2013}. Especially, latency as a performance indicator is critical for drone-cells. The flow- and cloud-based networking are promising approaches to overcome these challenges~\cite{bradai_cellular_2015, sezer_are_2013, zhou_toward_2014, yazici}. 

Flow-based networking requires advancements, such as developing new routing protocols, interfaces, and applications. The major difficulties associated with the cloud are centralizing the distributed data, providing security, determining the level of sharing while satisfying the regulations, and providing the power required for processing massive amounts of data~\cite{demestichas_5g_2013, zhou_toward_2014}. In this sense, real-time collection and processing of the data required to manage a drone's operation (e.g. tackling turbulence, avoiding collisions, tracking user mobility) is infeasible. Therefore, DMF is unlikely to alleviate the need for drones with high levels of autonomy~\cite[Ch. 70]{handbook}, but DMF can provide guidelines, as demonstrated in the following section.

\section{A Case Study: 3-D Placement of a drone-cell via DMF}
\label{sec:virt}
Efficient placement is a critical and challenging issue for drone-cells. In this section, we propose an objective for DMF, meeting various demands simultaneously. Then, we numerically illustrate the benefit of using DMF by comparing the results with the efficient 3-D placement\footnote{3-D placement concept is introduced in~\cite{bor_2016} because the probability of having LOS connection increases with increasing altitude, and, at the same time, path loss increases due to increased distance. Therefore, an optimum altitude is sought after, as well as an optimal area to cover in the horizontal domain.} method proposed in~\cite{bor_2016}, and show that DMF can split costs among MVNOs without detracting from the network benefit in a multi-tenancy model.
%
%
%
%

Let us consider that a drone-cell, managed via DMF, is used to assist a terrestrial HetNet with the following considerations:
\begin{itemize}
\item \textbf{Congestion release in RAN:} A set of users, $\mathbb{U}$, cannot be served by the BS because of the congestion. The objective is to serve as many users from the set $\mathbb{U}$ as possible by the drone-cell. Let $u_i$ denote a binary variable indicating if the $i_{th}$ user in $\mathbb{U}$ is served by a drone-cell with orthogonal resources. Note that $\mathbb{U}$ is determined by MVNOs based on connection characteristics of each user~\cite{bradai_cellular_2015} (Table~\ref{tab:info}).
\item \textbf{Multi-tenancy:} An InP owns the drone-cell and sends it to the congested macrocell according to the intelligence provided by the cloud (Fig. 2). This network structure allows sharing the drone-cell's resources, if desired, to maximize the revenue and reduce the OPEX. Assuming all users provide the same revenue (as in~\cite{bor_2016}), the number of users associated with an MVNO and served by the drone-cell can be a measure of the revenue provided to that MVNO. Hence, the objective becomes maximizing the number of served users, as well as forcing the drone-cell to serve the target number of users of each MVNO. Then, if the total number of MVNOs in the macrocell is $\mathrm{J}$, a $\mathrm{J}\times 1$ vector $\mathbf{v}$ can be calculated, such that its $j^{th}$ element, $v_j$, denotes the ideal number of MVNO\textsubscript{j}'s users to be served by the drone-cell. Also, the cloud must store the vector $\mathbf{u}$ containing the indicator variables, $u_i$, and the matrix $\mathbf{S}$, which  denotes the user-MVNO associations. $S(i,j)\in\{0,1\}$ indicates if user $i$ belongs to MVNO $j$, which can be known from the subscriber information in the cloud (Table~\ref{tab:info}). Note that $\mathbf{v}$ is derived by cloud computing, based on several factors, such as agreements between the InP and MVNOs, pricing, user mobility, requested contents, and the scenario (Table~\ref{tab:info}, Fig.~\ref{fig:model}). 

\item \textbf{Green wireless communications:} $\mathbf{\Lambda}$ represents the energy cost of users. Hence, the drone-cell can be placed close to the energy critical users, such as sensor-type devices, or those at the blind spots (\encircle{7} in Fig.~\ref{fig:multi_tier}). Device-type information is collected by the MVNO (Table~\ref{tab:info}).

\item \textbf{Content-centric placement/Congestion release at the backhaul:} $\kappa_i$ indicates if the user $i$ requests a popular and costly (e.g., in terms of bandwidth or price) content, $\kappa$, which is cached in the drone-cell. Hence, the placement can be adjusted according to the content requirements of the users. Note that decisions about which contents to be delivered depends on the short- and long-term data collected by SPs on usage, user habits, and so on (Table~\ref{tab:info} and Fig.~\ref{fig:model}).  
\end{itemize}

Then, a comprehensive placement problem can be written as  
\begin{align*}
\label{eq:big_pic}
&\underset{\mathbf{p},\{u_i\}}{\text{max}}
\quad \omega_1\sum_{i \in U}u_{i} + \omega_2\|\mathbf{Su} - \mathbf{v} \| + \omega_3\|\mathbf{Su} - \mathbf{\Lambda}\| + \omega_4u_{i}\kappa_i \nonumber \\
& \text{s.t.}\qquad Q(\mathbf{p}, u_i) \leq \mathbf{q_i},
\ \ \forall i = 1,...,|\mathbb{U}|,\nonumber\\
& \ \ \qquad \quad \mathbf{p} \in \mathbb{P},\nonumber\\
& \ \qquad \quad \sum_{i \in\mathbb{U}}u_i R_i \leq C, \ \ \forall i = 1,...,|\mathbb{U}|,\numberthis\\
& \ \ \qquad \quad u_{i} \in \{0,1\},
\ \ \quad \forall i = 1,...,|\mathbb{U}|,\nonumber\\
\end{align*}
where $|\cdot|$, and $\|\cdot\|$ represent the cardinality of a set and vector norm operation, respectively; $\omega$ represents the weight of each benefit; $\mathbf{p}$ denotes the location of the drone-cell in 3-D space; $Q(\mathbf{p}, u_i)$, $q_i$, and $R_i$ denote the QoS delivered to the $i^{th}$ user from the drone-cell at location $\mathbf{p}$, the minimum tolerable service quality, and the required resources to serve the $i^{th}$ user, respectively. $C$ represents the capacity of the drone-cell and $\mathbb{P}$ denotes the set of allowable locations for placing the drone-cell, such as the allowed distance from the buildings according to regulations, or the positions with LOS links to the backhaul/fronhaul node. Note that the weights among the benefits, $\omega_i$, can be determined based on their importance to the owner of the drone-cells. Similarly, determining $\omega_i$, $\mathbf{v}$, and $\kappa_i$, based on their importance to the owner of the drone-cells, are interesting problems themselves.

The generic problem in~\eqref{eq:big_pic} is mathematically formulated in~\cite{bor_2016} by assuming $\omega_1 = 1$, and the rest of the weights are 0. We numerically compare the efficiency of DMF in this scenario by assuming multi-tenancy with 1 InP and 2 MVNOs serving the congested macrocell in an urban environment. In order to focus on the effect of multi-tenancy, we assume $w_1 = w_2 = 1$, and $w_3 = w_4 = 0$. There are 24 users that cannot be served by the terrestrial HetNet. They are distributed uniformly and arbitrarily subscribed to one of the two available MVNOs. The QoS requirement for all users is the minimum signal-to-noise ratio (100 dB maximum tolerable path loss). Also, MVNOs are identical, e.g. in terms of their agreements with InP, user priorities, and QoS requirements. Therefore, $v_1 = v_2= 12$, which is in favour of providing an equal amount of service to each MVNO. Hence, they can share the cost of the drone-cell equivalently. 

\begin{figure}[!t]
\centering
\includegraphics[width=0.45\textwidth, height=0.45\textwidth]{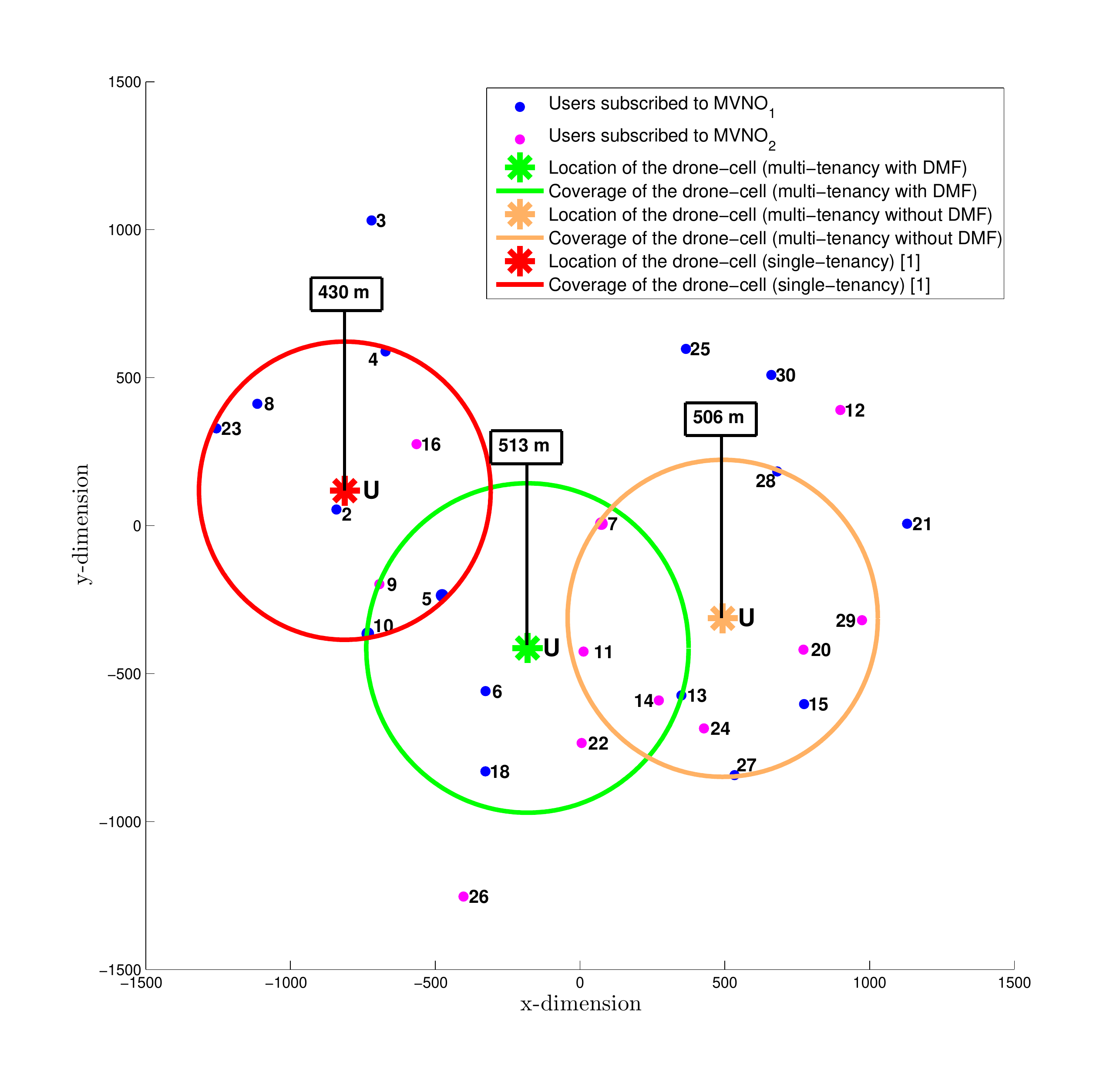}
\caption{Effect of different policies on 3-D placement of a drone-BS: Air-to-ground channel model in~\cite{bor_2016} relates the size of a  drone-cell to the altitude of the drone-BS. Therefore, both horizontal and vertical coordinates of a drone-BS must be determined simultaneously. Hence, an efficient 3-D placement algorithm is proposed to find the optimal altitude, as well as an optimal area to cover in the horizontal domain~\cite{bor_2016}. In this study, 3-D placement of a drone-cell is improved over~\cite{bor_2016} to regulate multi-tenancy by DMF, which ensures equivalent service is provided to both MVNOs. In the case of single-tenancy, only users subscribed to MVNO\textsubscript{1} are served by the drone-BS (blue users 2, 4, 5, 8, 10, 23). Note that not only single- or multi-tenancy (red vs. green and orange circles), but also regulating the service among MVNOs changes the placement (green vs. orange circles).}
\label{fig:cell}
\end{figure}

Fig.~\ref{fig:cell} shows how the placement of a drone-cell changes with respect to policies, namely, single-tenancy and multi-tenancy with and without DMF. The circular areas indicate the coverage of the drone-cell, and enclosed users are served by the drone-BS, i.e., their QoS requirements are satisfied. However, users of MVNO\textsubscript{2} (users 9 and 16) are not served in the red drone-cell due to single-tenancy policy. In other words, only 6 blue users (2, 4, 5, 8, 10, 23) are served. On the other hand, 10 users are enclosed in both green and orange drone-cells with multi-tenancy. In the orange drone-cell representing the placement without DMF, 4 users belong to MVNO\textsubscript{1} and 6 users belong to MVNO\textsubscript{2}. Hence, the resources of the drone-BS are not equally distributed as suggested by the cloud. That may reduce the benefit of the network, e.g., MVNO\textsubscript{1} may reject the drone-BS's services. However, when DMF is considered, 5 users of each MVNO are served in the green drone-cell. At the same time, there is no compromise in the network's benefit, since the total number of served users remains the same in both multi-tenancy scenarios. 

\begin{figure}[!t]
\centering
\includegraphics[width=\columnwidth]{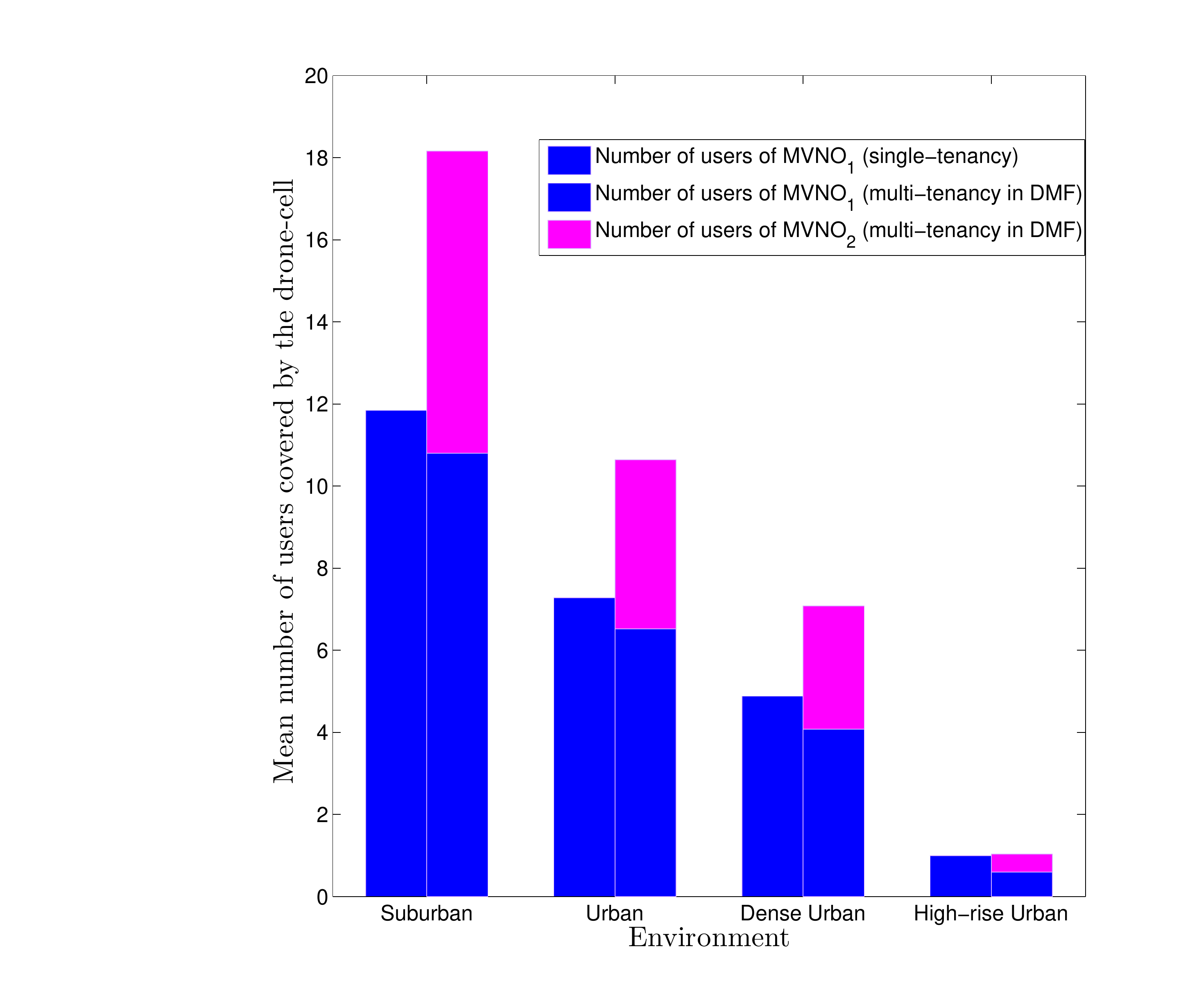}
\caption{Mean number of users covered by the drone-cell with 3-D placement in different environments is calculated after 100 Monte Carlo simulations. Drone-BS only serves the users of MVNO\textsubscript{1} for the ``single-tenancy" sceario. When drone-BS serves both MVNOs for the ``multi-tenancy" scenario of, DMF ensures fair user association.}
\label{fig:number}
\end{figure}

In order to clarify the advantage of DMF, we consider two network configurations. In the first one, we assume that the drone-cell only serves the users of MVNO\textsubscript{1} (e.g., blue dots in Fig.~\ref{fig:cell}, red drone-cell). In the second, we assume that both user groups exist. A comparison of the two cases is provided in Fig.~\ref{fig:number}, where 30 idle users in four different environments are randomly distributed~\cite{bor_2016}, and the results of 100 Monte Carlo simulations are averaged. It shows that MVNO\textsubscript{1} serves almost the same number of users (1-2 users less in each case) when it shares the drone-cell with MVNO\textsubscript{2}. In turn, the drone-cell's cost can be reduced by a factor of two. Moreover, the total number of served users increases (approximately 1.5 times), which means that more congestion is released from the network. 

Although it has remained implicit due to the limitations of the article, the number of covered users can also indicate the amount of injected capacity, enhanced coverage, and reduced re-transmission time in a congested scenario. Moreover, we have demonstrated the 3-D placement of one drone-cell, although, multi-tier drone-cell networks require additional considerations, such as inter-cell interference, cell density, cooperation of drone-cells, and green networking. Therefore, collecting data to define the problem in~(1), and then analyzing it efficiently requires a holistic and centralized cellular network, rather than the existing distributed one. The better drone-cells are managed, the more the advantages of their flexibility can be exploited.

\section{Conclusion}

The ultra-dense small cell deployment has attracted significant attention in recent years as an advanced radio access architecture to cope with extreme traffic demands. However, the fact that such extreme demands can often be sporadic and hard to predict in space and time renders an ultra-dense deployment (which will end up being under-utilized most of the time) highly inefficient and even prohibitive from a cost perspective. The multi-tier drone-cell network envisioned in this article is a new radio access paradigm that enables bringing the supply of wireless networks to where the demand is in space and time.
%

We discussed the potential advantages and challenges of integrating drone-cells in future wireless networks with a holistic and detailed approach from the mechanics of drone-BSs to potential applications of advanced networking technologies. Considering the fact that wireless networks are mainly designed for the mobility of the users but not the BSs, and that the drone-cell operations can be highly complex, we proposed a novel DMF (drone management framework) for an efficient operation. We demonstrated the proposed DMF and its benefits via a case study, where drone-cells are utilized in wireless networks with multi-tenancy.

%

\ifCLASSOPTIONcaptionsoff
  \newpage
\fi



%

\bibliographystyle{IEEEtran}
\bibliography{magazine}

%
\begin{IEEEbiographynophoto}{Irem Bor Yaliniz}(irembor@sce.carleton.ca) received B.Sc. and M.Sc. degrees in Electrical and Electronics Engineering from Bilkent University, Turkey in 2009 and 2012 respectively. She worked in Aselsan, which is a leading defence company, where she was a design engineer for physical and data layer embedded coding of Professional Radio Systems. She received scholarships through the Engage grant of the Natural Sciences and Engineering Research Council of Canada (NSERC) in 2014, and the Queen Elizabeth II Scholarship in Science and Technology in 2015.
\end{IEEEbiographynophoto}
\begin{IEEEbiographynophoto}{Halim Yanikomeroglu}
(halim@sce.carleton.ca) is a full professor in the
Department of Systems and Computer Engineering at Carleton University, Ottawa, Canada. His research interests cover many aspects of wireless technologies with special emphasis on cellular networks. He has co-authored more than 80 IEEE journal papers on wireless technologies. His collaborative research with industry has resulted in about 25 patents (granted and applied). He is a Distinguished Lecturer for the IEEE Communications Society and a Distinguished Speaker for the IEEE Vehicular Technology Society.
\end{IEEEbiographynophoto}
\end{document}